\documentclass[11pt,draftcls,journal,onecolumn]{IEEEtran}

\usepackage{graphicx, epsfig}

\begin{document}

\title{Joint Equalization and Decoding for Nonlinear Two-Dimensional Intersymbol Interference Channels with Application to Optical Storage}

\author{Naveen Singla and Joseph A. O'Sullivan}

\maketitle

\begin{abstract}
An algorithm that performs joint equalization and decoding for nonlinear two-dimensional intersymbol interference channels is presented. The algorithm performs sum-product message-passing on a factor graph that represents the underlying system. The two-dimensional optical storage (TWODOS) technology is an example of a system with nonlinear two-dimensional intersymbol interference. Simulations for the nonlinear channel model of TWODOS show significant improvement in performance over uncoded performance. Noise tolerance thresholds for the algorithm for the TWODOS channel, computed using density evolution, are also presented and accurately predict the limiting performance of the algorithm as the codeword length increases.
\end{abstract}


\section{Introduction}{\label{intro}}

Two-dimensional (2D) intersymbol interference (ISI) channels have received a lot of attention lately. This is mainly due to the fact that research focus in storage is shifting towards developing a two-dimensional storage paradigm. Although conventional recording media like magnetic hard disks and DVDs use planar storage, the second dimension is utilized only loosely. Significant increase in storage density can be obtained by moving towards truly two-dimensional storage. Patterned magnetic media~\cite{white}, holographic storage~\cite{ashley}, and two-dimensional optical storage~\cite{coene},~\cite{coene1} (TWODOS) are examples of upcoming technologies that use a two-dimensional storage paradigm and promise terabit storage density at high data rates. Due to the two-dimensional nature of storage, these advanced storage technologies have 2D ISI during the readback process. Conventional methods like partial response maximum-likelihood decoding, that have proved very successful for one-dimensional ISI channels, do not extend to two dimensions. This motivates the need for new methods to combat 2D ISI. Besides advanced storage technologies, multi-user communication scenarios, like cellular communication, also have situations where 2D ISI is prevalent. 


Several detection (or equalization) algorithms for 2D ISI channels have been proposed [5-11]. Some of these algorithms are based on linear minimum mean-square-error (MMSE) equalization; the MMSE equalization is followed by (soft or hard) thresholding and may iterate between equalization and thresholding. Others are based on multi-track versions of the Viterbi or BCJR algorithm accompanied by decision feedback. Singla \emph{et al.}, [11-14] have proposed joint equalization and decoding schemes for 2D ISI channels and have shown the benefit of using error control coding in conjunction with detection. More often than not, the ISI is modeled as a linear filter. Although a good starting point, the linearity assumption doesn't hold in general. TWODOS is an example of a system where the ISI is nonlinear. 

TWODOS is, potentially, the next generation optical storage technology with projected storage capacity twice that of the blu-ray disk and with ten times faster data access rates~\cite{coene},~\cite{coene1}. As in conventional optical disk recording, bits in the TWODOS model are written on the disk in spiral tracks. However, instead of having a single row of bit cells, each track consists of a number of bit rows stacked together making TWODOS a truly two-dimensional storage paradigm. Successive tracks on the disk are separated by a guard band which consists of one empty bit row. In addition, the bit cells are hexagonal; this allows 15 percent higher packing density than rectangular bit cells leading to even higher storage capacity. As in conventional optical disk recording, a 0/1 is represented by the absence/presence of a pit on the disk surface. A scalar diffraction model proposed by Coene~\cite{coene} for optical recording is used to model the readback signal from the disk. Under this model the readback intensity from the disk has linear and bilinear contributions from the stored data bits.

Various detection schemes for TWODOS have been proposed [15-17]. These schemes, with the exception of Immink \emph{et al.},~\cite{twodos}, use two-dimensional partial response equalization to obtain a linear channel model for the ISI. Then, equalization methods, like MMSE equalization, are used for detection on this linearized channel model. Since partial response equalization leads to noise correlation there is an inherent loss associated with these schemes. Thus, it is prudent to search for decoding schemes that avoid partial response equalization and are designed taking into account the nonlinear structure of the ISI. Immink \emph{et al.},~\cite{twodos} propose using a stripe-wise Viterbi detector that is designed for the nonlinear ISI. Chugg \emph{et al.}, also proposed equalization schemes for nonlinear 2D ISI channels~\cite{chugg1},~\cite{chugg2}. However, neither of the aforementioned schemes employed error control coding.

In this paper, a low-complexity scheme for joint equalization and decoding for nonlinear 2D ISI channels is presented. The scheme was first proposed for linear 2D ISI channels~\cite{singla1} and has been appropriately modified for the nonlinear channel. This scheme, called the full graph scheme, performs sum-product message-passing on a joint graph that represents the error control code and the nonlinear 2D ISI channel. Low-density parity-check (LDPC) codes~\cite{mackay} are used for error correction. Simulations for the nonlinear channel model of TWODOS demonstrate the potential of using the full graph scheme. Significant improvement in performance is observed over uncoded performance. Noise tolerance thresholds are calculated for regular LDPC codes of different rates and sum-product decoding for the nonlinear 2D ISI channel. We note that others have also proposed message-passing based schemes for joint equalization and decoding for a wide variety of channels like the fading channel~\cite{stark} and partial response channels~\cite{kurkoski},~\cite{kavcic}.

The paper is organized as follows. The system model and the channel model for TWODOS is described in Section~\ref{channelmodel}. The full graph message-passing algorithm and its performance for TWODOS are presented in Section~\ref{FGMP}. The density evolution algorithm and the noise tolerance thresholds are presented in Section~\ref{denevol}. Section~\ref{conclusion} concludes the paper.

\section{System model}{\label{channelmodel}}

\noindent{The system is modeled as a discrete-time communication system;}

\begin{equation}
r(i,j)=h(\{x(k,l):(k,l){\in}{\mathcal{N}_{ij}}\})+w(i,j),
\label{eq:one}
\end{equation}

\noindent{where $r(i,j)$ are the data received at the output of the channel; $x(k,l)$ are the channel inputs, obtained by encoding the user data with an error control code; $w(i,j)$ are samples of additive white Gaussian noise (AWGN) with zero mean and variance $\sigma^{2}$ and are assumed to be independent of $x(k,l)$; $\mathcal{N}_{ij}$ is the set of indices of all the bits that interfere with $x(i,j)$ during readback, including $x(i,j)$; and $h(\cdot)$ is the function that encapsulates the nonlinear 2D ISI. The user data and the encoded data are assumed to be binary. LDPC codes are used for error correction.

For TWODOS, a scalar diffraction model proposed by Coene~\cite{coene} for optical recording is used to model the readback signal. Using the model, the readback signal (optical intensity) from the disk can be written as

\begin{equation}
r(i,j)=1-\sum_{(k,l)}c_{ij}(k,l)x(k,l)+\sum_{(k,l){\neq}(m,n)}d_{ij}(k,l;m,n)x(k,l)x(m,n)+w(i,j),
\label{eq:two}
\end{equation}

\noindent{where $c_{ij}(k,l)$ and $d_{ij}(k,l;m,n)$ are the linear and nonlinear ISI coefficients, respectively. These coefficients depend on the parameters of the optical system, such as the wavelength of the laser, numerical aperture of the readback lens and geometry of the recording (pit and track dimensions). The extent of the interference is limited by the spot size of the read laser. Typically, this restricts the interference to nearest neighbors only. As shown in~\cite{coene} this assumption is quite accurate. 

Using a nearest neighbor interference model, the signal intensity in~(\ref{eq:two}) depends on the data bit stored in the central bit cell and the 6 neighboring bit cells. If it is assumed that two configurations with the same central bit and same number of nonzero neighbors have identical signal values then the signal intensity takes on 14 values corresponding to the 14 different configurations. As shown in~\cite{coene}, this symmetry assumption is a good approximation. Fig.~\ref{fig:hex} shows four of these 14 configurations.  

\begin{figure}[ht]
\centering
\scalebox{0.7}{\includegraphics[angle=270,clip]{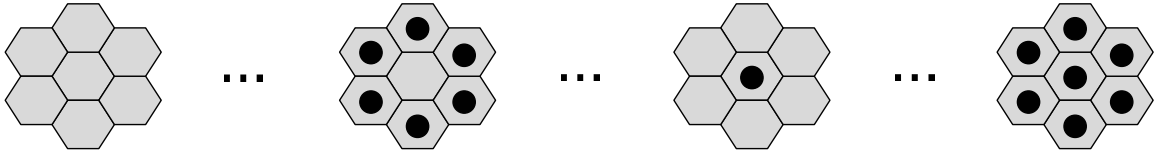}}
  \caption{Four of the possible 14 nearest neighbor configurations for TWODOS. The dark circles in the cells depict a pit corresponding to a stored 1. Absence of a pit indicates a stored 0. The pits cover only about half the area of the hexagonal bit cells. This is done to reduce signal folding~\cite{coene}.}
  \label{fig:hex}
\end{figure}

Table~\ref{tbl:siglevel} lists the signal levels for the 14 different configurations for one choice of pit dimensions and laser spot size. This table is reproduced from~\cite{coene}. Looking at the table, the nonlinearity of the ISI is quite apparent; the signal level does not change linearly as the number of nonzero neighbors increases and the range when the central bit is a 0 is greater than when the central bit is 1. 

\begin{table}[ht]
\begin{center}
\caption{Signal levels for TWODOS recording using nearest neighbors interference. (Reproduced from~\cite{coene})}
\label{tbl:siglevel}
\begin{tabular}{|c|c|c|}
\hline
Nonzero       & Central            & Central             \\
neighbors (n) & bit=0 $(s_{n_0})$  & bit=1 $(s_{n_1})$   \\
\hline
0 & 0.95 & 0.50 \\
\hline
1 & 0.80 & 0.35 \\
\hline
2 & 0.70 & 0.30 \\
\hline
3 & 0.55 & 0.20 \\
\hline
4 & 0.45 & 0.15 \\
\hline
5 & 0.35 & 0.10 \\
\hline
6 & 0.25 & 0.05 \\
\hline
\end{tabular}
\end{center}
\end{table}

\section{Full Graph Message-Passing Algorithm}{\label{FGMP}}

The decoder uses the sum-product algorithm to compute the maximum \emph{a posteriori} probability estimate of the codeword given the channel output. The graph on which the algorithm operates represents the following factorization:

{\setlength\arraycolsep{2pt}
\begin{eqnarray}
p({\bf{X}}{\mid}{\bf{R}}) & {\propto} & p({\bf{R}}{\mid}{\bf{X}})P({\bf{X}}) \nonumber \\
& \propto & \prod_{(i,j)}^{}p(r(i,j){\mid}\{x(k,l):(k,l){\in}{\mathcal{N}}(i,j)\})P({\bf{X}}),
\label{eq:three}
\end{eqnarray}}

\noindent{where $\mathbf{X}$ and $\mathbf{R}$ are matrices representing the channel input and channel output, respectively. $\mathcal{N}(i,j)$ is as defined previously. $P(\mathbf{X})$ is the probability that $\mathbf{X}$ is a codeword of the LDPC code being used. This probability can be represented graphically via the Tanner graph of the LDPC code~\cite{ksch}. $p({\bf{R}}{\mid}{\bf{X}})$ represents the ISI channel.}

Fig.~\ref{fig:threelevel} shows an illustration of this factor graph where nearest neighbor interference is assumed. This ``full graph'' has three types of nodes: variable nodes, check nodes, and measured data nodes corresponding to the codeword bits, the parity-check equations, and the observed data symbols, respectively. The upper two levels in the full graph represent the LDPC code bipartite graph showing how the codeword bits are connected to the check nodes via the LDPC code parity-check matrix. The lower two levels represent the channel ISI graph showing how the ISI induces dependencies between the codeword bits. For clarity, connections for only one measured data node are shown in the figure.

\begin{figure}[ht]
\centering
\scalebox{0.6}{\includegraphics[angle=270,clip]{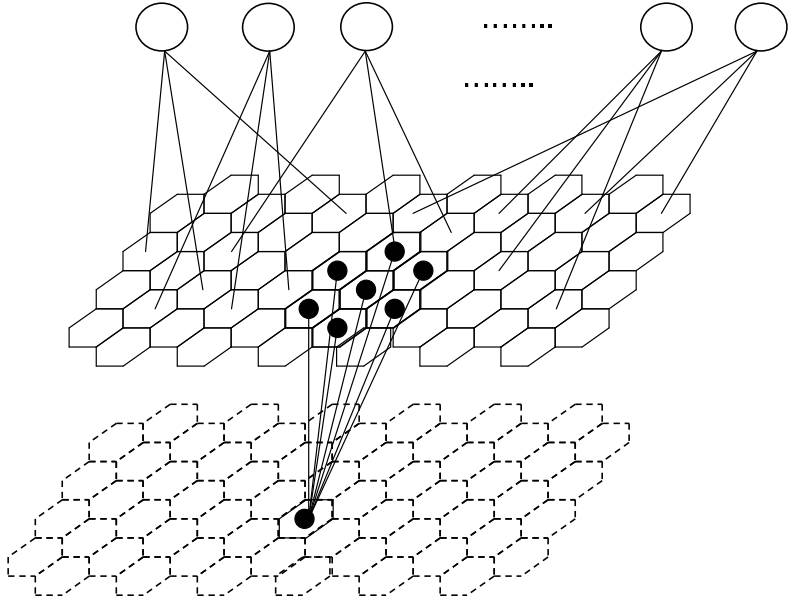}}
  \caption{The full graph depicting the factorization of~(\ref{eq:three}).}
  \label{fig:threelevel}
\end{figure}

Message-passing on this full graph is performed using the following schedule of messages: variable nodes to check nodes, check nodes to variable nodes, variable nodes to measured data nodes and finally measured data nodes to variable nodes. Following is a brief description of how the messages are updated for each of the aforementioned four steps for the sum-product algorithm. The messages in the update equations are probabilities.

\noindent{{\bf{Variable-to-check messages:}} The message from a variable node $x$ to a check node $c$ at the $l$th iteration is calculated using the messages passed to $x$ from its neighboring check and measured data nodes at the $(l-1)$th iteration.}

\begin{eqnarray}
\mu_{x{\rightarrow}c}^{(l)}(0) & = & \alpha\prod_{r{\in}N_r(x)}\mu_{r{\rightarrow}x}^{(l-1)}(0)\prod_{c'{\in}N_c(x){\setminus}c}\mu_{c'{\rightarrow}x}^{(l-1)}(0) \nonumber \\
\mu_{x{\rightarrow}c}^{(l)}(1) & = & \alpha\prod_{r{\in}N_r(x)}\mu_{r{\rightarrow}x}^{(l-1)}(1)\prod_{c'{\in}N_c(x){\setminus}c}\mu_{c'{\rightarrow}x}^{(l-1)}(1),
\label{eq:four}
\end{eqnarray}

\noindent{where $\mu_{x{\rightarrow}c}^{(l)}({\cdot})$, $\mu_{r{\rightarrow}x}^{(l)}({\cdot})$, and $\mu_{c{\rightarrow}x}^{(l)}({\cdot})$ are, respectively, the variable-to-check, measured data-to-variable, and check-to-variable messages at the $l$th iteration. $N_c(x)$ and $N_r(x)$ are the neighboring check and measured data nodes of $x$ respectively, and $\alpha$ is a normalizing constant.}

\noindent{{\bf{Check-to-variable messages:}} At the $l$th iteration the check-to-variable messages are calculated using the variable-to-check messages at the $l$th iteration and the sum-product rule,}

\begin{eqnarray}
\mu_{c{\rightarrow}x}^{(l)}(0) & = & \sum_{\tilde{\mathbf{x}}_{c}}P(c{\mid}x=0,\tilde{\mathbf{x}}_{c})\prod_{x'{\in}N(c){\setminus}x}\mu_{x'{\rightarrow}c}^{(l)}(x') \nonumber \\
\mu_{c{\rightarrow}x}^{(l)}(1) & = & \sum_{\tilde{\mathbf{x}}_{c}}P(c{\mid}x=1,\tilde{\mathbf{x}}_{c})\prod_{x'{\in}N(c){\setminus}x}\mu_{x'{\rightarrow}c}^{(l)}(x')
\label{eq:five}
\end{eqnarray}

\noindent{where $N(c)$ are the variable nodes connected to check node $c$; $\tilde{\mathbf{x}}_{c}$ are length $|N(c)|-1$ binary tuples. The conditional probabilities above are 0 or 1 depending on whether the parity-check constraint at node $c$ is satisfied or not.}

\noindent{{\bf{Variable-to-measured data messages:}} The variable-to-measured data messages at the $l$th iteration are calculated using the messages received at the variable nodes from the check nodes at the $l$th iteration and from the measured data nodes at the $(l-1)$th iteration,}

\begin{eqnarray}
\mu_{x{\rightarrow}r}^{(l)}(0) & = & \alpha\prod_{c{\in}N_c(x)}\mu_{c{\rightarrow}x}^{(l)}(0)\prod_{r'{\in}N_r(x){\setminus}r}\mu_{r'{\rightarrow}x}^{(l-1)}(0) \nonumber \\
\mu_{x{\rightarrow}r}^{(l)}(1) & = & \alpha\prod_{c{\in}N_c(x)}\mu_{c{\rightarrow}x}^{(l)}(1)\prod_{r'{\in}N_r(x){\setminus}r}\mu_{r'{\rightarrow}x}^{(l-1)}(1).
\label{eq:six}
\end{eqnarray}

\noindent{{\bf{Measured data-to-variable messages:}} To complete one iteration, the measured data-to-variable messages at the $l$th iteration are computed using the variable-to-measured data messages at the $l$th iteration and the sum-product rule,}

\begin{eqnarray}
\mu_{r{\rightarrow}x}^{(l)}(0) & = & \alpha\sum_{\tilde{\mathbf{x}}_{r}}p(r{\mid}x=0,\tilde{\mathbf{x}}_{r})\mu_{x'{\rightarrow}r}^{(l)}(x') \nonumber \\
\mu_{r{\rightarrow}x}^{(l)}(1) & = & \alpha\sum_{\tilde{\mathbf{x}}_{r}}p(r{\mid}x=1,\tilde{\mathbf{x}}_{r})\mu_{x'{\rightarrow}r}^{(l)}(x')
\label{eq:seven} 
\end{eqnarray}

\noindent{$N(r)$ is the set of all variable nodes connected to measured data node $r$; $\tilde{\mathbf{x}}_{r}$ are length $|N(r)|-1$ binary tuples. The conditional probabilities above are values of a Gaussian probability density function. The mean of this probability density function is determined by the signal levels given in Table~\ref{tbl:siglevel} and the variance is equal to the noise variance $\sigma^{2}$.} 

After this step the ``pseudo-posterior'' probabilities are calculated using the messages from the check nodes and the measured data nodes,

\begin{eqnarray}
q_{x}^{(l)}(0) & = & \alpha\prod_{r{\in}N_r(x)}\mu_{r{\rightarrow}x}^{(l)}(0)\prod_{c{\in}N_c(x)}\mu_{c{\rightarrow}x}^{(l)}(0) \nonumber \\
q_{x}^{(l)}(1) & = & \alpha\prod_{r{\in}N_r(x)}\mu_{r{\rightarrow}x}^{(l)}(1)\prod_{c{\in}N_c(x)}\mu_{c{\rightarrow}x}^{(l)}(1). 
\label{eq:eight}
\end{eqnarray}

\noindent{The codeword estimate is obtained by setting $\hat{x}$ to 1 if $q_{x}^{(l)}(1)>q_{x}^{(l)}(0)$ and 0 otherwise. The decoding stops if the decoder converges to a codeword or a maximum number of iterations are exhausted. The complexity of the full graph algorithm is linear in the LDPC code block length and quadratic in the size of the interference neighborhood.}

Results of using full graph message-passing for the TWODOS channel model of~(\ref{eq:two}) and Table~\ref{tbl:siglevel} are shown in Fig.~\ref{fig:twodosresults}. The LDPC code used is a block length 10000, regular (3,30) code~\cite{rich1}. This high-rate code is chosen so as to add only a small number of redundant parity bits. The SNR is defined as the average energy in the signal divided by the noise power;

\begin{equation}
SNR=10{\log}_{10}\frac{\sum_{n=0}^{6}{6 \choose n}(s_{n_0}^{2}+s_{n_1}^{2})}{2^7{\cdot}(2R{\sigma}^2)},
\label{eq:nine}
\end{equation}

\noindent{where $s_{n_0}$($s_{n_1}$) is the signal level given that the central bit is a 0(1) and has $n$ nonzero neighbors and $R$ is the LDPC code rate. From right to left the solid curves in Fig.~\ref{fig:twodosresults} correspond to the performance of the full graph for a maximum of 1, 2, 3, 4, and 5 iterations. The number of iterations is kept low so as to reduce decoding delay. The dashed curves correspond to the uncoded performance which is the performance when the full graph algorithm iterates only between equations~(\ref{eq:six}) and~(\ref{eq:seven}) ignoring the LDPC code. This then gives a low-complexity detection scheme for the nonlinear ISI channel. As the curves show, the improvement in performance is quite significant. At a bit error rate of $10^{-5}$ the coded performance after 5 iterations is about 8 dB better than the uncoded performance after 10 iterations. Also, the performance of the full graph algorithm after 5 iterations is about 8 dB better (at a bit error rate of $10^{-6}$) than the performance reported in~\cite{twodos}, where a stripe-wise Viterbi detector is used for the same ISI.}

\begin{figure}
\centering
\scalebox{0.75}{\includegraphics{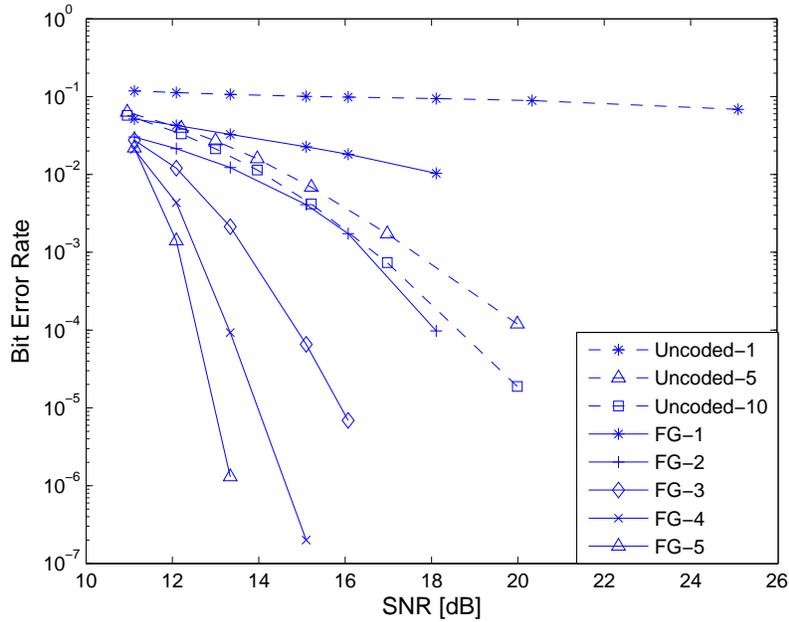}}
  \caption{Results of using the full graph message-passing algorithm for the TWODOS recording model given by~(\ref{eq:two}) and Table~\ref{tbl:siglevel}. The LDPC code used is a block length 10000 regular (3,30) code.}
  \label{fig:twodosresults}
\end{figure}

\section{Density Evolution and Threshold Computation}{\label{denevol}}

For many channels and decoders of interest, LDPC codes exhibit a threshold phenomenon~\cite{rich1}; there exists a critical value of the channel parameter (noise tolerance threshold, say $\delta^{*}$) such that an arbitrarily small bit error probability can be achieved if the noise level ($\delta$) is smaller than $\delta^{*}$ and the code length is long enough. On the other hand, for $\delta>\delta^{*}$ the probability of bit error is larger than a positive constant. Richardson and Urbanke~\cite{rich1} developed an algorithm called \emph{density evolution} for iteratively calculating message densities, enabling the determination of the aforementioned threshold. Kav\v{c}i\'{c} \emph{et al.},~\cite{kavcic} extended the work of Richardson and Urbanke to compute noise tolerance thresholds for one-dimensional ISI channels.

Using a density evolution similar to that proposed by Kav\v{c}i\'{c} \emph{et al.},~\cite{kavcic} noise tolerance thresholds are computed for the full graph algorithm and the nonlinear TWODOS channel. Following is a brief description of the algorithm. Density evolution tracks the ``evolution'' of the probability density function (pdf) of correct (or incorrect) messages passed on the graph as the iterations progress. Density evolution is described more conveniently using log-likelihood ratios (LLR) for messages instead of probabilities. The LLR for the variable-to-check messages at the $l$th iteration is defined as $L_{x{\rightarrow}c}^{(l)}=\log\frac{\mu_{x{\rightarrow}c}^{(l)}(0)}{\mu_{x{\rightarrow}c}^{(l)}(1)}$. The LLRs $L_{c{\rightarrow}x}, L_{x{\rightarrow}r},$ and $L_{r{\rightarrow}x}$ are defined analogously. Using LLRs gives an equivalent representation of the full graph message-passing algorithm. Equations~(\ref{eq:four})-(\ref{eq:seven}) can be rewritten as

\begin{eqnarray}
\label{eq:LLR1}
L_{x{\rightarrow}c}^{(l)} & = & \sum_{r{\in}N_{r}(x)}L_{r{\rightarrow}x}^{(l-1)}+\sum_{c'{\in}N_{c}(x){\setminus}c}L_{c'{\rightarrow}x}^{(l-1)} \\
\label{eq:LLR2}
\tanh(\frac{L_{c{\rightarrow}x}^{(l)}}{2}) & = & (-1)^{c}\prod_{x'{\in}N(c){\setminus}x}\tanh(\frac{L_{x'{\rightarrow}c}^{(l)}}{2}) \\
\label{eq:LLR3}
L_{x{\rightarrow}r}^{(l)} & = & \sum_{c{\in}N_{c}(x)}L_{c{\rightarrow}x}^{(l)}+\sum_{r'{\in}N_{r}(x){\setminus}r}L_{r'{\rightarrow}x}^{(l-1)} \\
\label{eq:LLR4}
L_{r{\rightarrow}x}^{(l)} & = & F(L_{x'{\rightarrow}r}^{(l)};x'{\in}N(r){\setminus}x),
\end{eqnarray}

\noindent{The update for measured data-to-variable messages has no closed form when represented using LLRs. The ``tanh'' rule~\cite{ksch} cannot be applied to~(\ref{eq:seven}) since the measured data nodes are not binary-valued. Equation~(\ref{eq:LLR4}) represents the measured data-to-variable message update via a function $F$ which performs the appropriate computation.} 

Let $f_{v}^{(l)}(x)$ be the pdf of the correct message from a variable node to a check node at the $l$th round of message-passing. This pdf is evolved through~(\ref{eq:LLR1})-(\ref{eq:LLR4}). The evolution of the density functions through~(\ref{eq:LLR1}) and~(\ref{eq:LLR3}) are simple convolutions and can be implemented efficiently using the fast Fourier transform. Density evolution for~(\ref{eq:LLR2}) can be implemented using the change in measure described by Richardson and Urbanke in~\cite{rich1} or more efficiently by using a table-lookup as explained by Chung \emph{et al.}, in~\cite{chung}. For density evolution through~(\ref{eq:LLR4}) Monte Carlo simulations are used; message-passing is performed on the channel graph using a long block length and the pdf of the outgoing messages from the measured data nodes is approximated by using the histogram of computed messages.

After evolving $f_{v}^{(l)}(x)$ through~(\ref{eq:LLR1})-(\ref{eq:LLR4}), $f_{v}^{(l+1)}(x)$ is obtained. Using $f_{v}^{(l+1)}(x)$, the error probability at the $(l+1)$th iteration, $p_{e}^{(l+1)}$, can be computed as

\begin{equation}
p_{e}^{(l+1)}=\int_{-\infty}^{0} f_{v}^{(l+1)}(x)dx.
\label{eq:ten}
\end{equation}

The noise tolerance threshold, $\delta^{*}$, can be calculated as the supremum of all $\delta$ for which the error probability goes to zero as the iterations progress. Table~\ref{tbl:thresholds} shows the computed thresholds for regular LDPC codes of different rates. The $(3,\infty)$ code refers to the case when no coding is used. The noise tolerance threshold is the variance of the AWGN. The SNR is calculated as in~(\ref{eq:nine}) except that the rate of the code is not taken into account. 

\begin{table}[ht]
\begin{center}
\caption{Noise tolerance thresholds for the TWODOS channel.}
\label{tbl:thresholds}
\vspace{0.1in}
\begin{tabular}{|c|c|c|c|}
\hline
LDPC & Code & Threshold        & Threshold   \\
Code & Rate & $\sigma_{*}^{2}$ & SNR [dB]    \\
\hline
(3,3)   & 0.000 & 0.0670 & 1.7270    \\
\hline
(3,4)   & 0.250 & 0.0436 & 3.5929    \\
\hline
(3,5)   & 0.400 & 0.0283 & 5.4699    \\
\hline
(3,6)   & 0.500 & 0.0215 & 6.6633    \\
\hline
(3,9)   & 0.667 & 0.0140 & 8.5264    \\
\hline
(3,12)  & 0.750 & 0.0117 & 9.3059    \\
\hline
(3,15)  & 0.800 & 0.0103 & 9.8593    \\
\hline
(3,30)  & 0.900 & 0.0061 & 12.1344   \\
\hline
(3,60)  & 0.950 & 0.0035 & 14.5470   \\
\hline
(3,90)  & 0.967 & 0.0030 & 15.2165   \\
\hline
(3,120) & 0.975 & 0.0027 & 15.6741   \\
\hline
(3,150) & 0.980 & 0.0025 & 16.0083   \\
\hline
(3,$\infty$) & 1.000 & 0.0018 & 17.4342 \\
\hline
\end{tabular}
\end{center}
\end{table}

In proving the existence of thresholds for memoryless channels the crucial innovation of Richardson and Urbanke in~\cite{rich1} were the ``concentration results.'' These results state that as the block length tends to infinity the performance of the LDPC decoder on random graphs converges to its expected behavior and that the expected behavior can be determined from the corresponding cycle-free behavior. Kav\v{c}i\'{c} \emph{et al.},~\cite{kavcic} extended these concentration results to one-dimensional ISI channels by using LDPC coset codes to circumvent the complications arising out of the input-dependent memory.

For 2D ISI channels the concentration results do not hold since the channel graph has short cycles even in the limit of infinitely long block length. Hence existence of thresholds cannot be proved using the concentration analysis. However, our simulations suggest that for the TWODOS channel the full graph algorithm respects the thresholds computed using density evolution. These results are shown in Fig.~\ref{fig:denevolresults} for three block lengths: 10000, 65536, and 262144 and using a regular (3,30) LDPC code. The results show that very low bit error rates are obtained only when the noise variance is smaller than the threshold. Although this does not prove the existence of a threshold, it suggests that the noise tolerance thresholds of Table~\ref{tbl:thresholds} are upper bounds on the performance of the full graph algorithm. Besides that, the thresholds also serve as a design parameter; given a system with a specified SNR it is sufficient to pick an LDPC code having a smaller threshold SNR thereby ensuring that the bit-error rate can be made arbitrarily small as the block length increases.

\begin{figure}
\centering
\scalebox{0.75}{\includegraphics{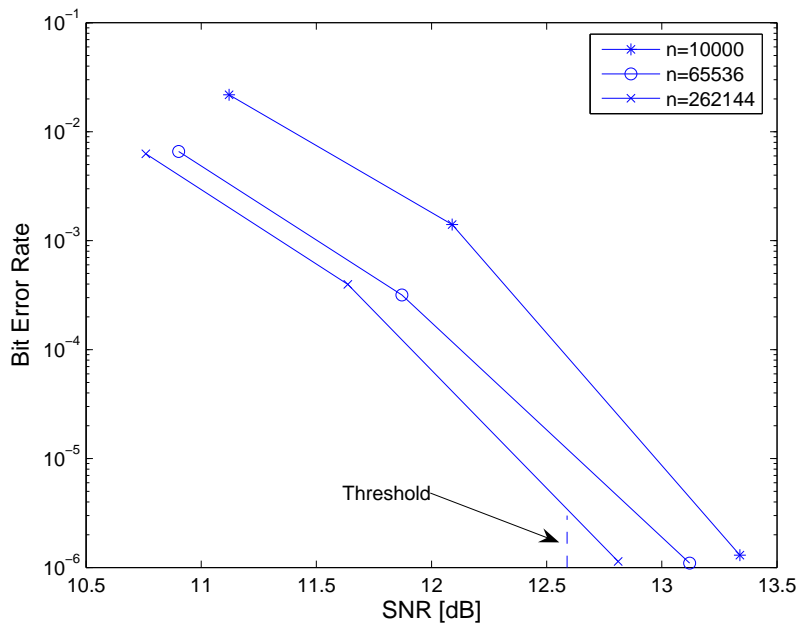}}
  \caption{Results of using the full graph message-passing algorithm for the TWODOS recording model using different length regular (3,30) LDPC codes.}
  \label{fig:denevolresults}
\end{figure}

\section{Conclusions}{\label{conclusion}}

A message-passing based scheme for joint equalization and decoding for nonlinear two-dimensional intersymbol interference channels has been proposed. The scheme, called the full graph algorithm, performs sum-product message-passing on a joint graph of the error correction code and the channel. The complexity of the full graph algorithm is linear in the block length of the error correction code and quadratic in the size of interference neighborhood. The performance of the algorithm is studied for the two-dimensional optical storage paradigm. Simulations for the nonlinear channel model of TWODOS show significant improvement over uncoded performance. The performance is about 6 to 8 dB better than that reported in by Immink \emph{et al.}, ~\cite{twodos} for the same intersymbol interference. A detection scheme for nonlinear ISI channels has also been proposed. This scheme performs sum-product message-passing on the graph corresponding to the nonlinear channel ISI. Using density evolution noise tolerance thresholds for the full graph algorithm are also computed and are shown to accurately predict the performance of the algorithm as the block length gets large.

\end{document}